\documentclass{article}
\usepackage{spconf,amsmath,graphicx}
\usepackage{multirow}
\usepackage{tikz}
\usepackage{booktabs}
\usepackage{csquotes}
\usepackage{multirow}
\usepackage{caption,subcaption}
\usepackage{enumitem}
\usepackage{pbox}
\usepackage{pifont}
\usepackage{enumitem}
\usepackage{mathtools}

\title{Foundation Model Assisted Automatic Speech Emotion Recognition: Transcribing, Annotating, and Augmenting}
\name{Tiantian Feng$^{1}$, Shrikanth Narayanan$^{1}$}
\address{$^{1}$Signal Analysis and Interpretation Laboratory \\University of Southern California, Los Angeles, USA}

\begin{document}
\ninept
\maketitle
\begin{abstract}
Significant advances are being made in speech emotion recognition (SER) using deep learning models. Nonetheless, training SER systems remains challenging, requiring both time and costly resources. Like many other machine learning tasks, acquiring datasets for SER requires substantial data annotation efforts, including transcription and labeling. These annotation processes present challenges when attempting to scale up conventional SER systems. Recent developments in foundational models have had a tremendous impact, giving rise to applications such as ChatGPT. These models have enhanced human-computer interactions including bringing unique possibilities for streamlining data collection in fields like SER. In this research, we explore the use of foundational models to assist in automating SER from transcription and annotation to augmentation. Our study demonstrates that these models can generate transcriptions to enhance the performance of SER systems that rely solely on speech data. Furthermore, we note that annotating emotions from transcribed speech remains a challenging task. However, combining outputs from multiple LLMs enhances the quality of annotations. Lastly, our findings suggest the feasibility of augmenting existing speech emotion datasets by annotating unlabeled speech samples.
\end{abstract}


\section{Introduction}
\label{section:intro}


Speech emotion recognition (SER) has benefited considerably from using large-scale pre-trained speech models \cite{baevski2020wav2vec,schneider2019wav2vec,chen2022wavlm,radford2022robust}, offering substantial performance improvements over conventional SER systems that primarily depend on low-level acoustic descriptors (e.g., speech prosody and spectral information). These advances in emotion recognition open up opportunities for widespread applications in healthcare and virtual assistants, transforming our ways of connecting, engaging, and interacting with the world. However, success in deploying SER models in real-world applications requires the acquisition of high-quality annotations to speech samples, which is often expensive, time-consuming, and privacy-unfriendly. 


One typical labeling step in SER datasets involves transcribing the speech content. For example, IEMOCAP \cite{busso2008iemocap}, one of the most popular SER testbeds, had obtained the professional transcriptions of the audio dialogues using a commercial service. Such a process often requires training transcribers on transcription guidelines, creating considerable R\&D costs. The advent of Amazon's Mechanical Turk\cite{marge2010using} (MTurk) had substantially increased the efficiency of transcribing services by providing the marketplace for human workers to perform such tasks for pay. However, it still demands many MTurk hours to transcribe the audio conversations, leading to significant costs. In addition, MTurk may not be a viable option when the data collection poses significant privacy risks and must be annotated in-house, which is a standard practice mandated by Institutional Review Boards (IRBs) involving sensitive human subject data \cite{feng2022review}.

Furthermore, SER dataset often requires emotion labeling. A standard emotion labeling process involves instructing multiple human annotators to assess the emotional content of the speech sample in terms of emotional descriptors. Similar to transcribing, the emotion annotation procedure yields substantial costs in hiring multiple annotators to ensure authentic appraisal of a speech sample. Moreover, utilizing services such as MTurk for emotion annotation would raise notable privacy risks. Therefore, curating the SER dataset remains a challenging task, particularly for institutions that encounter resource constraints and comply with strict regulatory guidelines.

The emergence of foundation models \cite{bommasani2021opportunities} delivered promising speech recognition and language reasoning performances, bringing unique opportunities to facilitate SER data curation. For example, Whisper \cite{radford2022robust} is designed for automatic speech recognition (ASR), trained on thousands of hours of audio data from the Internet. This model delivers remarkable zero-shot ASR performance, demonstrating its enormous potential for deployment as a transcription service. Along with the advancements in automatic transcription, large language models (LLMs) like GPT4 \cite{OpenAI2023GPT4TR} offer human-level text reasoning and comprehension capabilities, positioning them as candidates for reducing the involvement of human emotion annotation.

In this paper, we report comprehensive experiments on the use of foundation models in assisting curation of the speech emotion recognition dataset in transcribing, emotion annotation, and augmentation. Our study focuses on exploring modeling approaches that require a single V100-equivalent GPU, ensuring the ease of reproducibility. In summary, our contributions are listed as follows: 


\begin{figure*}
    \centering
    \includegraphics[width=0.9\linewidth]{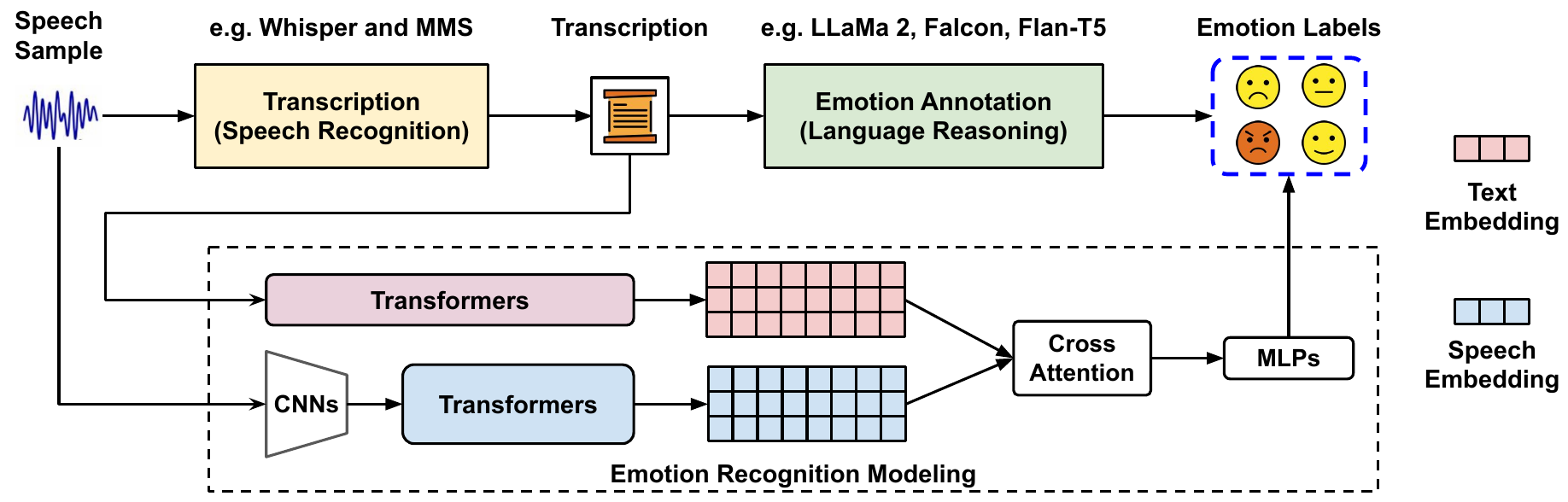}
    \vspace{-2mm}
    \caption{Our proposed foundation model-assisted automatic SER framework. The speech is first transcribed to text and is subsequently fed to LLMs to annotate categorical emotions. Our SER modeling framework involves a text and speech backbone to extract corresponding embeddings, which are then passed through a cross-attention layer to obtain the multimodal representations to predict emotion labels.}
    \label{fig:framework}
    \vspace{-3mm}
\end{figure*}

\vspace{-1mm}
\begin{itemize}[leftmargin=*]
    \item Our work represents one of the early studies on the use of the foundation model to assist SER dataset curation covering three critical factors: \textbf{transcribing}, \textbf{emotion annotation}, and \textbf{augmentation}.
    
    \vspace{-1mm}
    \item Our experiments study Whisper and MMS as transcribing annotators, where we find that existing foundation model systems provide transcriptions that are beneficial for SER training. 

    \vspace{-1mm}
    \item We investigate using multiple open-source LLMs as emotion annotators, revealing that emotion annotation remains challenging for LLMs. Moreover, combining limited human annotations with LLM output substantially improves the SER training.

    \vspace{-1mm}
    \item We explore data augmentation using the foundation model-assisted annotations, leading to increases in SER performance.
    
\end{itemize}

\vspace{-3mm}
\section{Related Works}

\subsection{Speech Recognition Models}

Self-supervised learning (SSL) is a rapidly emerging research area for speech representation learning. This learning approach enables the pre-trained speech models, which are then trained with labeled speech samples for speech-related tasks. One recent popular model in this category is the Massively Multilingual Speech (MMS) \cite{pratap2023scaling} model released by Meta, which is pre-trained on 491K hours of speech. In contrast, Whisper by OpenAI \cite{radford2022robust} adopts a weakly supervised learning approach, with objectives to perform tasks such as voice activity detection, language identification, and speech recognition. The training of this model is conducted using a dataset comprising 680k hours of labeled speech data. 

\vspace{-2mm}
\subsection{Large Language Models} 

Large language models like ChatGPT have demonstrated remarkable performance in language reasoning tasks. However, GPT4 or ChatGPT requires the user to upload the speech content to the remote server for prompting. This creates considerable privacy risks in sensitive settings and applications. Instead, we decided to explore foundation models that can operate on a single GPU, including LLaMa 2 families \cite{touvron2023llama}, Falcon families \cite{falcon}, and Flan-T5 XXL \cite{chung2022scaling}. We want to highlight that several prior works \cite{gong23c_interspeech, latif2023can} have investigated the ability of LLMs to annotate ground-truth transcriptions or ASR-generated transcription. However, most of these works consider conventional SER modeling architecture (e.g., ResNet-50). Moreover, they do not incorporate ASR-generated transcription in SER modeling and experiment with a limited set of LLMs.

\vspace{-2mm}
\section{Method}
\vspace{-2mm}
\subsection{Foundation Model Assisted Annotation}
Our automatic annotation framework is presented in Fig~\ref{fig:framework}. Given an unlabeled speech sample, we first propose to obtain the speech content using foundation speech recognition models. This work investigates two recent ASR models, Whisper-Large V2 and MMS, that offer the most competitive results. After obtaining the ASR-generated transcripts, we directly send them to the large language models. Our LLMs include LLaMa 2 families, Falcon families, and Flan-T5 XXL. The details about the foundation models used in this study and their approximate model size can be found in Table~\ref{table:foundation_models}. The obtained emotion labels and transcripts are used for SER training.

\begin{table}[t]
    \caption{Summary of foundation models used in this work.}
    \vspace{-2mm}
    \begin{tabular*}{\linewidth}{lccc}
        \toprule
        
        \textbf{Foundation Model} & 
        \textbf{Input} & 
        \textbf{Annotation} & 
        \textbf{\# Parameters}  \\ 

        \midrule
        \textbf{MMS-1B} & Speech & Transcription & ~1000M \\ 
        \textbf{Whisper Large V2} & Speech & Transcription & 1.550M \\
        \midrule
        \textbf{LLaMa 2-7B} & Text & Emotion & 7B \\ 
        \textbf{LLaMa 2-13B} & Text & Emotion & 13B \\
        \textbf{Falcon-7B} & Text & Emotion & 7B \\ 
        \textbf{Falcon-40B} & Text & Emotion & 40B \\
        \textbf{Flan-T5 XXL} & Text & Emotion & 11B \\

        \bottomrule
    \end{tabular*}
    \vspace{-4mm}
    \label{table:foundation_models}
\end{table}

\vspace{-1mm}
\subsection{A Bag of Tricks in Prompt Engineering}
We investigate and compare several tricks in prompt engineering.

\vspace{1mm}
\noindent \textbf{Base Prompt} Our prompt design is similar to \cite{latif2023can}, where instructing the LLMs to annotate the spoken utterance delivers decent zero-shot performance. In addition, we instruct the LLMs to choose emotions from five categories: neutral, sad, happy, angry, and other. This strategy constrains the LLMs to output more determined labels, and we introduce the option of "other" to filter out unconfident responses to include in SER modeling. In summary, our prompt template is: 

\noindent
\textcolor{blue} {
 What is the emotion of this utterance? "Everything is not working!"}
\textcolor{blue} {
\noindent Options: -neutral -sad -angry -happy -other ANSWER:
} \textbf{sad}

\vspace{1mm}
\noindent \textbf{Multiple-LLMs Agreement} It is known that relying on the response from one LLM could yield biased language reasoning \cite{kasneci2023chatgpt}. To mitigate this concern, we propose ensemble the output from multiple LLMs, collecting the wisdom from multiple reasoners. 

\vspace{1mm}
\noindent \textbf{LLMs + Human Feedback} One critical lesson we learned from prior research is that LLMs exhibit limited zero-shot capabilities in annotating emotions from speech. Consequently, we contend that human evaluation may remain essential. However, instead of relying on multiple human raters for a majority agreement, we propose that assessing the agreement between the LLM annotations and one human feedback is sufficient for quality control.


\vspace{-2mm}
\subsection{Emotion Recognition Modeling}
The complete model architecture is illustrated in Fig~\ref{fig:framework}. Our SER includes speech and text backbones to extract the corresponding embeddings. Specifically, we utilize Whisper-Small \cite{radford2022robust} and MMS-300M \cite{pratap2023scaling} as the speech backbone and Roberta as the text backbone. We intend not to experiment with Whisper-Large as the speech backbone as it requires prohibitively large GPU capacities for our setting. The output of backbone models is subsequently fed into weighted averaging layers to combine the hidden outputs from all encoder layers. The weighted output is then passed through a cross-attention layer to obtain the multimodal representation for SER.

\begin{table}[t]
    \caption{Summary of dataset statistics used in this work.}
    \small
    \vspace{-3mm}
    \begin{tabular*}{\linewidth}{lccccc}
        \toprule
        
        \multirow{1}{*}{\shortstack{\textbf{Datasets}}} & 
        \multirow{1}{*}{\textbf{Neutral}} & 
        \multirow{1}{*}{\shortstack{\textbf{Happy}}} &
        \multirow{1}{*}{\shortstack{\textbf{Sad}}} & 
        \multirow{1}{*}{\shortstack{\textbf{Angry}}} & 
        \multirow{1}{*}{\shortstack{\textbf{Total}}}  \\ 
         
        \midrule
        \textbf{IEMOCAP} & 1,708 & 1,636 & 1,084 & 1,103 & 5,531 \\ 
        \textbf{MELD} & 6,436 & 2,308 & 1,002 & 1,607 & 9045 \\ 
        \textbf{MSP-Improv} & 3,477 & 2,644 & 885 & 792 & 7,798 \\ 
        \textbf{MSP-Podcast} & 20,986 & 12,060 & 2,166 & 2,712 & 37,924 \\

        \bottomrule
    \end{tabular*}
    \vspace{-4mm}
\label{table:datasets}
\end{table}

\vspace{-2mm}
\section{Datasets}
Table~\ref{table:datasets} displays data statistics for the four datasets included in our work. Due to the existence of imbalanced label distribution within the dataset, we decided to keep the four most frequently presented emotions for all the datasets, as recommended in \cite{feng2022enhancing,feng2021privacy,chen2021exploring,pepino21_interspeech}. We acknowledge that this inclusion criterion trivializes the automatic emotion annotation, but it ensures fair comparisons when having multiple datasets with different emotions. The emotion annotation results reported in our experiments will likely decrease in practice.

\vspace{0.5mm}
\noindent \textbf{IEMOCAP} \cite{busso2008iemocap} contains multi-modal recordings of human interactions from 10 subjects evenly distributed between males and females. 

\vspace{0.5mm}
\noindent \textbf{Multimodal EmotionLines Dataset (MELD)} \cite{poria2018meld} contains more than 13000 utterances from the Friends TV series. Each utterance is labeled with seven emotions, -- Anger, Disgust, Sadness, Joy, Neutral, Surprise, and Fear. We map the Joy to happy emotion and keep Anger, Sadness, and Neutral in the experiments.

\vspace{0.5mm}
\noindent \textbf{MSP-Improv}~\cite{busso2016msp} corpus is developed to investigate naturalistic emotions elicited from improvised situations. The corpus comprises audio and visual data collected from 12 individuals, with an equal number of subjects from both male and female participants. 

\vspace{0.5mm}
\noindent \textbf{MSP-Podcast}~\cite{lotfian2017building} is collected from podcast recordings, with 610 speakers in the training, 30 in the development, and 50 in the test.

\vspace{-2mm}
\section{Experiment Details}
\vspace{-2mm}
\subsection{Foundation Model Assisted Annotation}

We apply MMS-300M and Whisper Large V2 to obtain the ASR output. Since LLMs with more than 10B parameters exceed most GPU memory capacities, we decided to load LLMs over 10B using float 16 instead of float 32. In addition, we load Falcon-40B with 8-bit. We use a temperature of 0.02 in all prompting experiments, as a lower temperature results in more deterministic output. We use the checkpoints of all foundation models from HuggingFace \cite{wolf-etal-2020-transformers}.

\vspace{-2mm}
\subsection{Emotion Recognition Modeling}
We apply a 5-fold and 6-fold evaluation on IEMOCAP and MSP-Improv datasets respectively, where each session is regarded as a unique test fold. In contrast, we use the standard splits for training, validation, and testing from the MELD and MSP-Podcast datasets. We use the RoBERTa \cite{liu2019roberta} model as the text backbone while we compare the speech backbones between MMS-300M and Whisper-Small. We choose MMS-300M along with MMS-1B ASR output and Whisper-Small along with Whisper Large V2 ASR output in SER modeling. Specifically, we set the batch size to 32, the learning rate to 0.0001, the max training epoch to 30, and truncated utterances to 15 seconds in baseline emotion recognition training. We use the ground-truth transcriptions in the test set for fair comparisons. We use the checkpoints of backbone models from HuggingFace \cite{wolf-etal-2020-transformers}.

\begin{table}
    \caption{SER performances using transcriptions.}
    \small
    \vspace{-3mm}
    \begin{tabular*}{\linewidth}{lcccc}
        \toprule
        
        \multirow{1}{*}{\shortstack{\textbf{Datasets}}} & 
        \multirow{1}{*}{\textbf{Input}} & 
        \multirow{1}{*}{\shortstack{\textbf{Transcription}}} &
        \multirow{1}{*}{\shortstack{\textbf{UAR(\%)}}} \\ 
         
        \midrule
        & Speech & - & 67.45 \\ 
        \textbf{IEMOCAP} & Speech+Text & Ground-truth & \textbf{73.87} \\ 
        & Speech+Text & Whisper-Large V2 & 71.78 \\ 
        \midrule
        
        & Speech & - & 48.55 \\ 
        \textbf{MELD} & Speech+Text & Ground-truth & \textbf{56.31} \\ 
        & Speech+Text & Whisper-Large V2 & 54.32 \\

        \midrule
        \multirow{2}{*}{\shortstack{\textbf{MSP-Improv}}} & Speech & - & 63.23 \\ 
        & Speech+Text & Whisper-Large V2 & \textbf{65.44} \\

        \midrule
        \multirow{2}{*}{\shortstack{\textbf{MSP-Podcast}}} & Speech & - & 60.82 \\ 
         & Speech+Text & Whisper-Large V2 & \textbf{63.19} \\

        \bottomrule
    \end{tabular*}
    \vspace{-4mm}
    \label{table:transcription_results}
\end{table}

\begin{figure}
    \centering
    \includegraphics[width=0.9\linewidth]{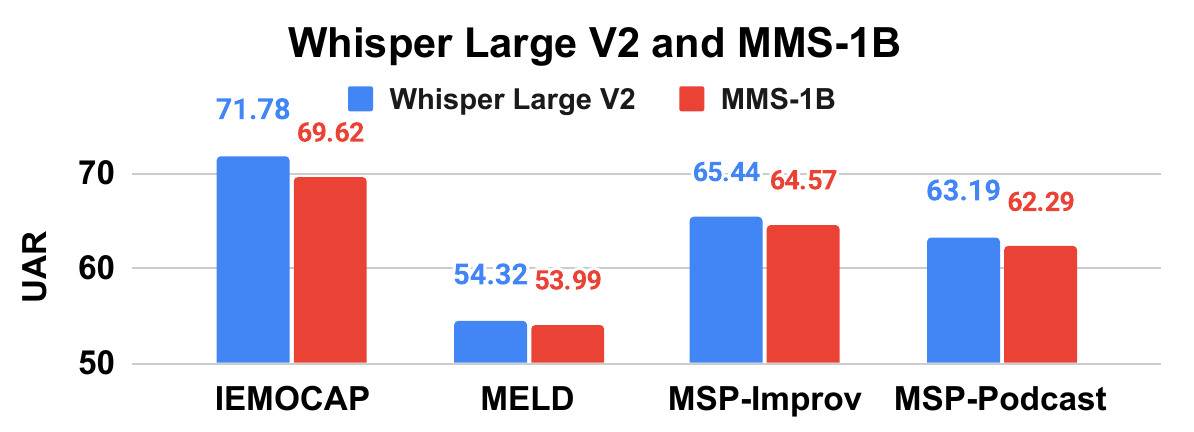}
    \vspace{-3mm}
    \caption{Comparisons between two foundation models in transcribing.}
    \label{fig:whisper_mms}
    \vspace{-4mm}
\end{figure}

\begin{table*}[t]
    \caption{SER (UAR) with emotion annotation from LLMs. The transcription is ASR output from Whisper Large V2. HF is human feedback.}
    \small
    \vspace{-3mm}
    \begin{tabular*}{\linewidth}{lccccccc}
        \toprule
        
        \multirow{1}{*}{\shortstack{\textbf{Datasets}}} & 
        \multirow{1}{*}{\textbf{Flan-T5 XXL}} & 
        \multirow{1}{*}{\shortstack{\textbf{LLaMa2-7B}}} &
        \multirow{1}{*}{\shortstack{\textbf{LLaMa2-13B}}} & 
        \multirow{1}{*}{\shortstack{\textbf{Falcon-7B}}} & 
        \multirow{1}{*}{\shortstack{\textbf{Falcon-40B}}} &
        \multirow{1}{*}{\shortstack{\textbf{Multi-LLMs}}} &
        \multirow{1}{*}{\shortstack{\textbf{Multi-LLMs+HF}}}  \\ 
         
        \midrule
        \textbf{IEMOCAP} & 49.60 & 43.87 & 46.29 & 43.68 & 51.16 & 51.60 & 60.19 \\ 

        \textbf{MELD} & 36.73 & 43.87 & 43.85 & 46.96 & 47.62 & 53.90 & NA \\ 
        
        \textbf{MSP-Improv} & 44.97 & 38.12 & 41.68 & 37.71 & 44.87 & 46.05 & 50.06 \\ 
        \textbf{MSP-Podcast} & 51.20 & 47.23 & 48.12 & 43.25 & 48.11 & 52.59 & 53.54 \\
        
        \bottomrule
    \end{tabular*}
    \vspace{-4mm}
    \label{table:llm_finetune}
\end{table*}

\begin{table}[t]
    \caption{WER (word error rate) in transcriptions. Processed transcripts consider only lowercase and remove punctuation.}
    \vspace{-3mm}
    \centering
    \begin{tabular*}{0.95\linewidth}{lcccc}
        \toprule
        
        \multirow{2}{*}{\shortstack{\textbf{Datasets}}} & 
        \multicolumn{2}{c}{\textbf{Whisper Large V2}} & 
        \multicolumn{2}{c}{\shortstack{\textbf{MMS-1B}}}  \\ 

        & Processed & Original 
        & Processed & Original \\ 
         
        \midrule
        \multirow{1}{*}{\shortstack{\textbf{IEMOCAP}}} & 12.21 & 24.84 & 26.76 & 51.46 \\ 

        \multirow{1}{*}{\shortstack{\textbf{MELD}}} & 37.87 & 46.23 & 55.78 & 71.28 \\ 

        \bottomrule
    \end{tabular*}
    \vspace{-3mm}
    \label{table:wer}
\end{table}

\vspace{-2mm}
\section{Transcription Results}

\vspace{-1mm}
\subsection{Does SER benefit from ASR using Foundation Model?}
This section compares the SER training using ASR-generated with ground-truth transcriptions (human transcriptions). As both MSP-Improv and MSP-Podcast datasets do not have transcriptions from human experts, we conduct SER training using only ASR output from selected foundation models. The results in Table~\ref{table:transcription_results} demonstrate that the foundation model provides transcriptions that lead to consistent performance increases compared to speech-only modeling. Moreover, we can identify that ASR-generated output delivers competitive SER performance compared to ground-truth transcripts. It is worth noting that our proposed SER training using ASR output from foundation models considerably outperforms conventional SER systems such as Dialogue RNN \cite{majumder2019dialoguernn} and CNN-attention \cite{peng2021efficient}.

\vspace{-2mm}
\subsection{Does SER vary with different Foundation Models?}

We further compare SER performance using ASR output between Whisper-Large V2 and MMS-1B, as illustrated in Figure~\ref{fig:whisper_mms}. The findings indicate that SER performance using ASR output provides consistent benefits to speech-only modeling approaches. However, we have noticed that SER with Whisper-Large V2 transcripts consistently outperforms using the MMS-1B transcripts. To identify the cause that may contribute to this performance difference, we inspect the WER of these two models on IEMOCAP and MELD datasets with ground-truth transcription shown in Tabel~\ref{table:wer}. The WER indicates that Whisper Large V2 yields better speech recognition than MMS-1B in our experimental datasets. However, we can observe that WER is still fairly large in both datasets, complying with the findings in \cite{li2023asr}. Therefore, we proceeded with the remaining experiments for LLM emotion annotation using Whisper Large V2.

\vspace{-1mm}
\section{Emotion Annotations}
\vspace{-1mm}
\subsection{How does base prompt perform compared to prior works?}
Table~\ref{table:llm_finetune} shows the SER training performance leveraging the emotion annotations using each individual LLM. Similar to previous work, we identify that LLMs struggle to provide correct emotion labels for SER training, leading to a 10-20\% decrease in performance compared to SER training using ground-truth emotion labels. Moreover, larger LLMs provide better emotion labels, with Falcon-40B yielding the best overall emotion annotations for SER training. 

\vspace{-2mm}
\subsection{Can majority vote of multi-LLMs improve annotation?}

Based on the individual performance of emotional annotation shown in Table~\ref{table:llm_finetune}, we decide to apply the majority votes of emotion annotations from Flan-T5 XXL, LLaMa2-13B, and Falcon-40B as the emotion labels. The results indicate that aggregating majority votes from multi-LLMs enhances the quality of emotion annotation. However, this improvement is only marginal, leading to a 1-2$\%$ increase in SER performance. This observation suggests that relying on LLMs alone, even when considering input from multi-LLMs, yields unsatisfactory labels compared to conventional human labeling methods.

\vspace{-2mm}
\subsection{Would adding limited involvement of human annotation benefit emotion annotation?}

The last column in Table~\ref{table:llm_finetune} involves the performance of SER adding human feedback (HF) in the annotation process. As MELD does not provide individual annotator labels, we exclude this dataset in this experiment. It is obvious that integrating limited human feedback can lead to substantial improvement in SER training. Our hypothesis is that text modality may often provide ambiguous information in determining the emotion labels, thus LLMs are prone to give erroneous estimations of the expressed emotion given limited modalities. Limited inspections on audio samples with human annotators offer a disambiguation process that increases the label quality.

\vspace{-2mm}
\subsection{How different are emotion annotations using transcriptions between ground-truth and ASR output?}

Table~\ref{table:groundtruth} reveals the SER training comparisons using emotion labels inferred from ground truth and ASR transcriptions. We report results with datasets that include the ground truth transcriptions. Interestingly, results in Table~\ref{table:groundtruth} show that ASR transcriptions, even with fairly large WER, lead to comparable SER performance to ground truth transcriptions. Moreover, LLMs with HF consistently outperform LLMs-only annotation. In future studies, it is worth studying why erroneous ASR output can yield comparable emotion reasoning using clean ground-truth transcriptions.

\begin{table}[t]
    \caption{SER (UAR) comparisons with annotations using ground-truth and Whisper transcriptions. HF means human feedback.}
    \footnotesize
    \centering
    \vspace{-3mm}
    \begin{tabular*}{0.95\linewidth}{lccc}
        \toprule
        
        \multirow{1}{*}{\shortstack{\textbf{Datasets}}} & 
        \multirow{1}{*}{\textbf{Transcription}} & 
        \multirow{1}{*}{\shortstack{\textbf{Multi-LLMs}}} &
        \multirow{1}{*}{\shortstack{\textbf{LLMs+HF}}}  \\ 
         
        \midrule
        \multirow{2}{*}{\shortstack{\textbf{IEMOCAP}}} & Ground truth & 50.36 & 59.08 \\ 

        & Whisper Large V2 & 51.60 &  60.19  \\ 
        \midrule

        \multirow{2}{*}{\shortstack{\textbf{Meld}}} & Ground truth & 55.69 & N.A.  \\ 

        & Whisper Large V2 & 53.90 & N.A.  \\ 
        \bottomrule
    \end{tabular*}
    \vspace{-3mm}
    \label{table:groundtruth}
\end{table}

\begin{table}[t]
    \caption{SER performance with augmentation. $\uparrow$ indicates an increase in SER performance using augmentation.}
    \footnotesize
    \vspace{-2mm}
    \centering
    \begin{tabular*}{0.96\linewidth}{lccc}
        \toprule
        
        \multirow{1}{*}{\shortstack{\textbf{Datasets}}} & 
        \multirow{1}{*}{\textbf{Augmentation}} & 
        \multirow{1}{*}{\shortstack{\textbf{Multi-LLMs}}} &
        \multirow{1}{*}{\shortstack{\textbf{LLMs+Human}}}  \\ 
         
        \midrule
        \multirow{2}{*}{\shortstack{\textbf{IEMOCAP}}} & MELD & 72.60 $\uparrow$ & N.A.  \\ 

        & MSP-Podcast & 69.29 $\downarrow$ & $\mathbf{72.62}$ $\uparrow$  \\ 
        \midrule

        \multirow{2}{*}{\shortstack{\textbf{MSP-Improv}}} & MELD & 65.05 $\downarrow$ & N.A.  \\ 

        & MSP-Podcast & 64.31 $\downarrow$ & $\mathbf{66.68}$ $\uparrow$  \\ 
        \bottomrule
    \end{tabular*}
    \vspace{-4mm}
    \label{table:llm_augmentation}
\end{table}

\vspace{-2mm}
\section{Augmentation}
This section explores the ability to use our proposed automated labeling framework to augment an existing training dataset. We choose the multiple-LLMs agreement and LLMs with human feedback to provide emotion labels from ASR transcriptions, as these two approaches yield higher SER. We select MSP-Podcast and MELD as the augmentation datasets as these two datasets originate from Internet sources. This experiment setup is similar to the previous work in \cite{latif2023can}. The comparison aligns with the prior work \cite{latif2023can} that augmenting IEMOCAP data with MELD using multi-LLMs labeling improves the performance. However, this finding does not hold when the training data is MSP-Improv. Moreover, augmenting SER training with MSP-Podcast using multi-LLM labeling consistently decreases the SER performance. On the other hand, we discover that augmenting data using LLM labeling with even limited human feedback consistently improves the SER performance, highlighting the importance of human feedback in emotional reasoning.

\vspace{-2mm}
\section{Conclusion}
In this paper, we explore the use of the foundation model in assisting curation of the SER datasets in transcribing, emotion annotation, and augmentation. Our study focuses on exploring open-source models that require a single V100-equivalent GPU that is widely accessible. Our study demonstrates that foundational models can generate transcriptions to enhance the performance of SER systems that rely solely on speech data. However, WERs are fairly large. Furthermore, we observe that annotating emotions from transcribed speech remains a challenging task, even when combining outputs from multiple LLMs. Lastly, our findings suggest the feasibility of augmenting existing speech emotion datasets by annotating unlabeled speech samples using a two-stage annotation process that includes limited human feedback. In summary, our results highlight the importance of human-in-the-loop for annotating emotion labels from speech signals. Our future work would use multi-modal approaches to assist automatic emotion annotation instead of only LLMs.

\bibliographystyle{IEEEtran}
\bibliography{refs}

\end{document}